\documentclass{PoS}

\usepackage{amsfonts}
\usepackage{amsmath}
\usepackage{subfigure} 

\graphicspath{{./figs/}}
\newcommand{\TB}{\textrm{B}}
\newcommand{\TI}{\textrm{I}}
\newcommand{\TP}{\textrm{P}}
\newcommand{\TV}{\textrm{V}}
\newcommand{\TA}{\textrm{A}}
\newcommand{\TT}{\textrm{T}}

\title{
  $B_K$ with improved staggered fermions: analysis using SU(2) staggered
chiral perturbation theory
}

\ShortTitle{
  SU(2) analysis of $B_K$
}

\author{\speaker{Boram Yoon}, Taegil Bae, Yong-Chull Jang,
  Hyung-Jin Kim, Jangho Kim,
  Jongjeong Kim, Kwangwoo Kim, Weonjong Lee\\
  Lattice Gauge Theory Research Center, CTP, and FPRD, \\
  Department of Physics and Astronomy,
  Seoul National University, Seoul, 151-747, South Korea \\
  E-mail: \email{wlee@snu.ac.kr}}

\author{Chulwoo Jung \\
  Physics Department, Brookhaven National Laboratory,
  Upton, NY11973, USA \\
  E-mail: \email{chulwoo@bnl.gov}}

\author{Stephen R. Sharpe\\
  Physics Department, University of Washington, Seattle, WA 98195-1560 \\
  E-mail: \email{sharpe@phys.washington.edu}}

\abstract{ We report updated results for $B_K$ calculated using HYP-smeared
  staggered  fermions on the MILC asqtad $2+1$ flavor lattices. 
  We use four different lattice
  spacings ($a \approx$ 0.12, 0.09, 0.06 and 0.045 fm) to control the
  continuum extrapolation. We use SU(2)
  staggered chiral perturbation theory to do the data analysis. We
  find that $B_K(\text{NDR}, \mu=2 \text{ GeV}) = 0.526 \pm 0.007
  \pm 0.024$ and $\hat{B}_K = B_K(\text{RGI}) = 0.720 \pm 0.010 \pm
  0.033$. Here the first error is statistical and the second
  systematic.  The dominant source of error is that due to our use
  of a truncated (one-loop) matching factor.
}

\FullConference{
  The XXVIII International Symposium on Lattice Field Theory, Lattice2010\\
  June 14-19, 2010\\
  Villasimius, Italy
}

\begin{document}

\section{Introduction} 
This is the first of four proceedings providing
an update of our calculation of $B_K$ using improved staggered fermions
(HYP-smeared valence quarks on the MILC asqtad lattices). 
%
%
Here we present results using fits based on
SU(2) staggered chiral perturbation theory (SChPT), 
the method that leads to our most accurate results.
We focus on the progress made since last year's
lattice proceedings, Ref.~\cite{ref:wlee-2009-1}.
The other proceedings present, respectively, 
results using fits based on SU(3) SChPT~\cite{ref:wlee-2010-3},
a study of some sources of error~\cite{ref:wlee-2010-4},
and a more detailed look at results from the
ultrafine ensemble~\cite{ref:wlee-2010-5}.

Table~\ref{tab:milc-lat} shows the present status of our
running.
In the last year, we increased the number of measurements
on the C4 and S1 ensembles (by factors of 10 and $>2$, respectively),
and added two new ensembles: F2 and U1.
The F2 ensemble allows us to further check the sea-quark mass dependence,
while the U1 ensemble provides a fourth lattice spacing.
We have also written a long article,
Ref.~\cite{ref:wlee-2010-1}, in which we explain
both the SU(2) and SU(3) SChPT calculations leading to
our fitting forms, describe our fitting methods, and
present our full error budget.
The results presented in Ref.~\cite{ref:wlee-2010-1}
are intermediate between those presented at last year's
lattice conference~\cite{ref:wlee-2009-1}
and those presented here.
In particular, Ref.~\cite{ref:wlee-2010-1} does
not include results from the U1 ensemble.

\begin{table}[h!]
\begin{center}
\begin{tabular}{c | c | c | c | c | c | c}
\hline
$a$ (fm) & $am_l/am_s$ & geometry & ID & ens $\times$ meas 
& $B_K$($\mu=2$ GeV) & status \\
\hline
0.12 & 0.03/0.05  & $20^3 \times 64$ & C1 & $564 \times 1$ &  0.557(14) & old \\
0.12 & 0.02/0.05  & $20^3 \times 64$ & C2 & $486 \times 1$ &  0.569(16) & old \\
0.12 & 0.01/0.05  & $20^3 \times 64$ & C3 & $671 \times 9$ &  0.565(5)  & old \\
0.12 & 0.01/0.05  & $28^3 \times 64$ & C3-2 & $275 \times 8$ &  0.570(5)  & old \\
0.12 & 0.007/0.05 & $20^3 \times 64$ & C4 & $651 \times 10$ & 0.562(5)  & \texttt{update} \\
0.12 & 0.005/0.05 & $24^3 \times 64$ & C5 & $509 \times 1$ &  0.554(11) & old \\
\hline
0.09 & 0.0062/0.031 & $28^3 \times 96$ & F1 & $995 \times 1$ & 0.544(12) & old \\
0.09 & 0.0031/0.031 & $40^3 \times 96$ & F2 & $678 \times 1$ & 0.547(10) & \texttt{new} \\
\hline
0.06 & 0.0036/0.018 & $48^3 \times 144$ & S1 & $744 \times 2$ & 0.539(7) & \texttt{update} \\
\hline
0.045 & 0.0028/0.014 & $64^3 \times 192$ & U1 & $305 \times 1$ & 0.527(11) & \texttt{new} \\
\hline
\end{tabular}
\end{center}
\caption{MILC asqtad ensembles used to calculate $B_K$.
  $B_K(\text{NDR}, 2 \text{ GeV})$ is obtained using the SU(2) 
  analysis with 4X3Y-NNLO fits (discussed in the text).
}
\label{tab:milc-lat}
\end{table}

\section{SU(2) SChPT Analysis}
Our analysis makes essential use of 
SChPT~\cite{ref:wlee-1999-1,ref:bernard-2003-1}
in order to remove artifacts associated with taste breaking.
The application of SChPT to $B_K$ was worked out for
SU(3) SChPT in Ref.~\cite{ref:sharpe-2006-1}.
The extensions to our mixed-action set-up, and to
SU(2) ChPT, were presented in Ref.~\cite{ref:wlee-2010-1}.
We summarize the findings of the latter work in this section.

SU(2) ChPT treats the strange quark as heavy, and
expands in $m_\pi^2/m_K^2$, as well as in the usual ratio
$m_\pi^2/\Lambda_\chi^2$ (with $\Lambda_\chi\sim1\;$GeV)
\cite{su2chpt1,su2chpt2}.
It has been argued that this is a more reliable way
of extrapolating kaon properties to the physical
quark masses than using SU(3) ChPT, in which the strange
quark is treated as light.
Whether or not this is true in general, it turns out
that, for our application, there is an additional
important advantage of SU(2) ChPT.
This is that, at next-to-leading-order (NLO), the
SU(2) SChPT expression contains no low-energy coefficients
(LECs) arising from discretization or truncation errors.\footnote{%
Truncation errors arise because we match the four-fermion
operator appearing
in $B_K$ to continuum regularization using one-loop perturbation theory.}
This is not the case in SU(3) ChPT, and, as a result, the latter
gives rise to cumbersome fitting forms with many parameters.
In SU(2) SChPT, by contrast, one has the same number of
parameters at NLO as for a fermion discretization with
chiral symmetry, such as domain-wall fermions.
We find~\cite{ref:wlee-2010-1}
\begin{eqnarray}
  f_\text{th} &=& d_0 F_0 + d_1 \frac{X_P}{\Lambda_\chi^2}  + d_2
  \frac{X_P^2}{\Lambda_\chi^4} + d_3 \frac{L_P}{\Lambda_\chi^2}
\label{eq:fth}
\end{eqnarray}
where the chiral logs reside in the function
\begin{eqnarray}
  F_0 &=& 1 + \frac{1}{32\pi^2 f_\pi^2} \Big\{
  \ell(X_\TI) + (L_\TI-X_\TI)\tilde\ell(X_\TI)
  - 2 \langle \ell(X_\TB) \rangle \Big\}
\label{eq:F0}  \\
  \langle \ell(X_\TB)\rangle &=&
  \frac{1}{16} \Big[
    \ell(X_\TI) + \ell(X_\TP) + 4 \ell(X_\TV)
    +4 \ell(X_\TA) + 6 \ell(X_\TT) \Big]\,.
  \\
  \ell(X) &=& X \left[\log(X/\mu_{\rm DR}^2) \right]\,,
\quad  \tilde\ell(X) = - {d\ell(X)}/{dX}
  \label{eq:l}
\end{eqnarray}
Here $X_\TB$ ($L_\TB$) is the squared mass
of the valence (sea) pion with taste B, which we know
from our simulations or those of the MILC collaboration.
It is only through these masses that the taste-breaking
enters at NLO.
The coefficients $d_i$ have an unknown dependence 
on $r_s = m_s/\Lambda_{QCD}$, 
and, in addition, $d_0$ is a function of $a^2$
and $\alpha_s^2$.

We have included an analytic NNLO term in Eq.~\ref{eq:fth},
with coefficient $d_2$, and we do fits both without
and with this term, labeled respectively as NLO and NNLO.
The latter fits are not full NNLO fits, but rather are used
to gauge the errors arising from truncating ChPT.

\section{Fitting and Results}

Our lattice kaon is composed of valence (anti)quarks
with masses $m_x$ and $m_y$, which are to be
extrapolated to $m_d^{\rm phys}$ and $m_s^{\rm phys}$, respectively.
On each lattice we use 10 valence masses:
\[
am_x,\ am_y = a m_s \times n/10 
\qquad \text{ with } \qquad n = 1,2,3,\ldots,10\,,
\]
where $ a m_s$ is the strange sea-quark mass (which lies
fairly close to the physical strange mass).
For our standard SU(2) fitting
we choose the lowest 4 values for $am_x$
and the highest 3 values for $am_y$,
calling the resulting fits ``4X3Y''.
This means that we use only 12 out of the possible 55 mass
combinations, but by doing so we
maintain the SU(2) condition $m_x/m_y \ll 1$.

\begin{figure}[tbhp]
\centering
\includegraphics[width=0.49\textwidth]
{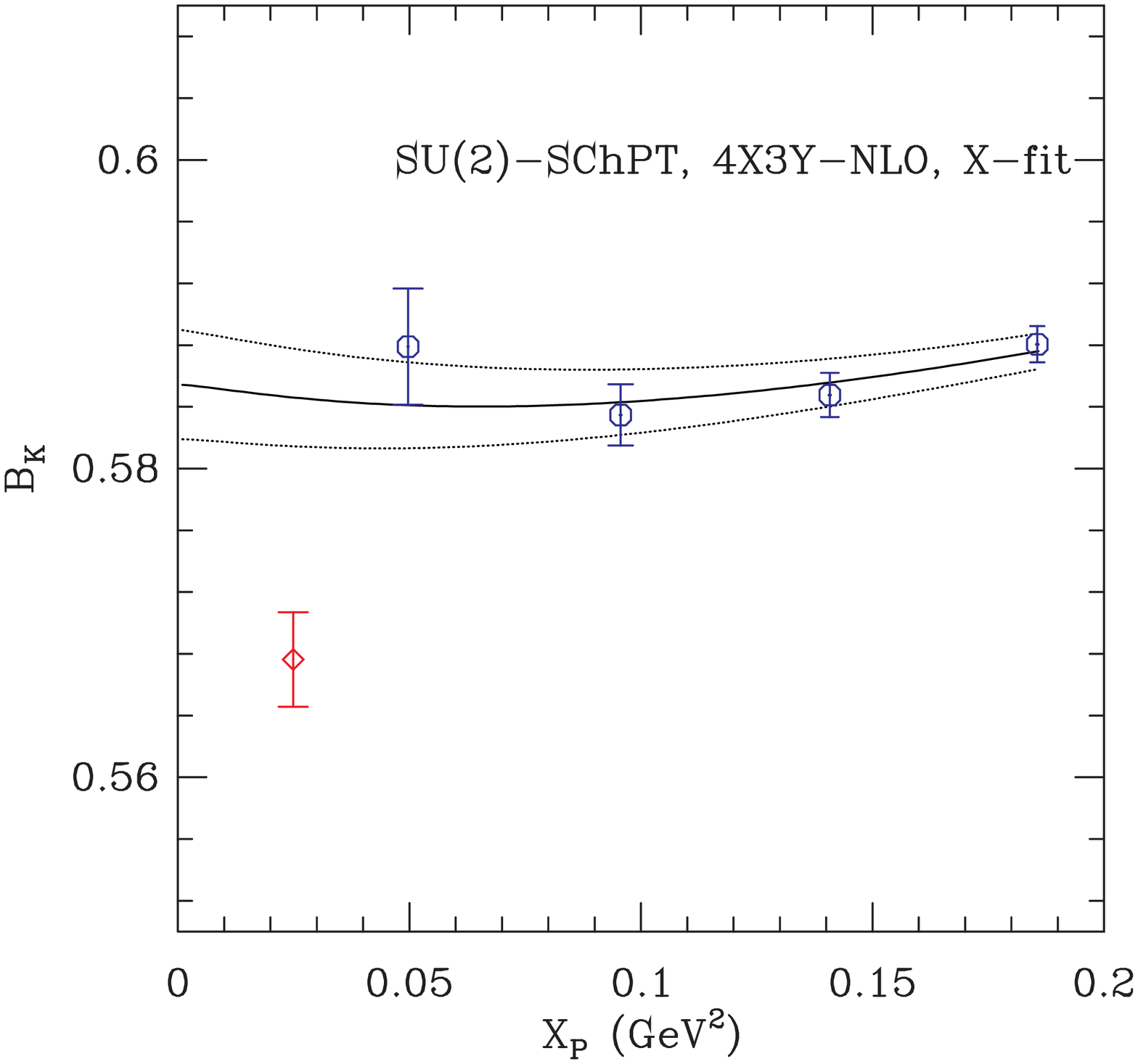}
\includegraphics[width=0.49\textwidth]
{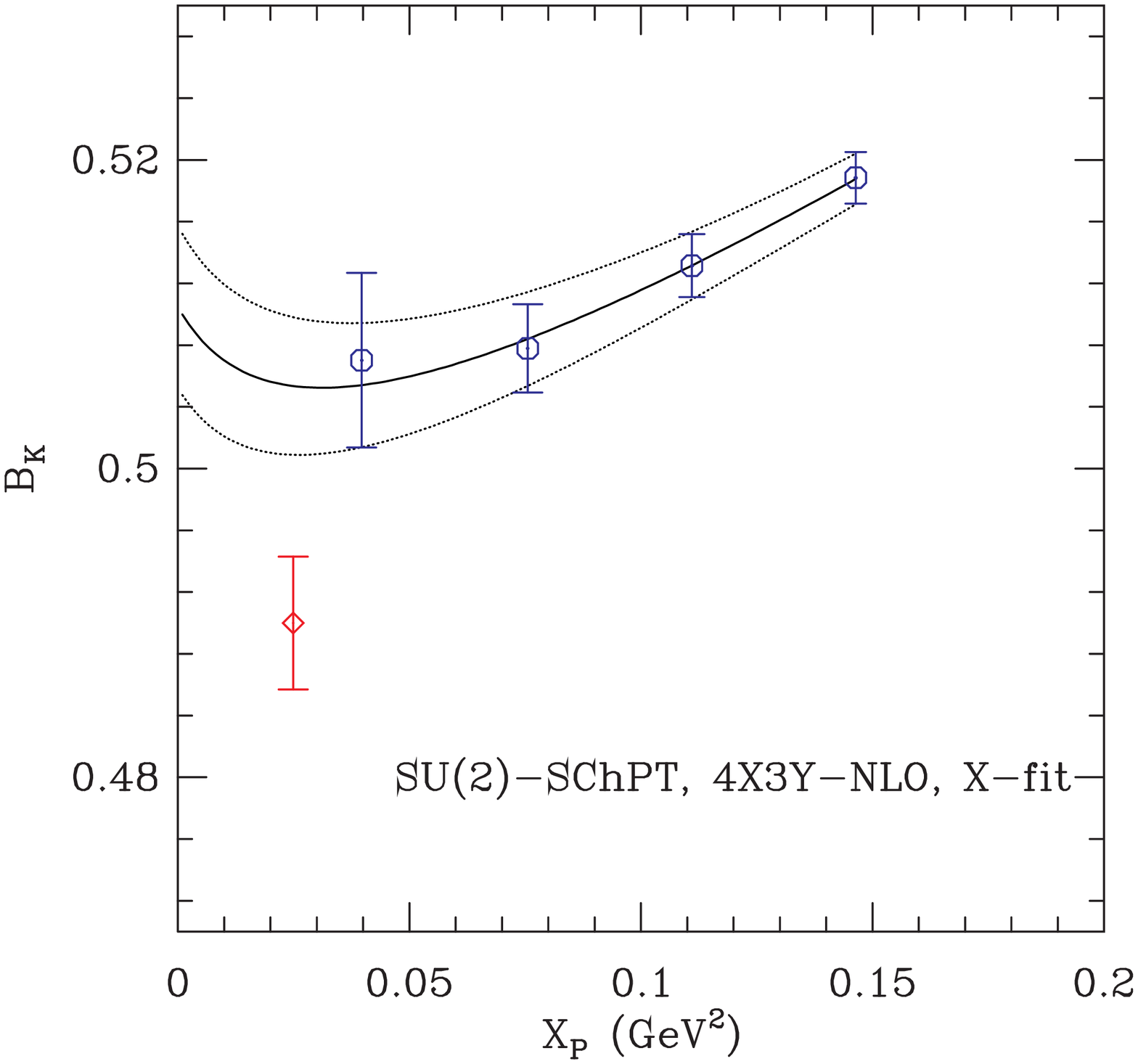}
\caption{ 4X-NLO fit of $B_K$ versus $X_P$ for the
  C3 (left) and S1 (right) ensembles. The red
  point gives the physical result, as discussed in
  the text.  Here, we fix $am_y = 0.05$ for C3 and
  $am_y = 0.018$ for S1. }
\label{fig:su2-4x3y-nlo}
\end{figure}

We do the fitting in two stages. First, for
fixed $m_y$, we fit the $m_x$ (or, equivalently, $X_P$) dependence 
to the form (\ref{eq:fth}). We call this the ``X-fit'',
examples of which are shown in Fig.~\ref{fig:su2-4x3y-nlo}.
This fit allows us to extrapolate $m_x\to m_d^{\rm phys}$
{\em and} to remove taste-breaking discretization and 
perturbative truncation errors by setting $X_I=X_B=X^{\rm phys}$
in Eq.~(\ref{eq:F0}). We can also partially correct for
the use of an unphysical $m_\ell$ by setting $L_I$ to
its physical value. The result of this procedure is shown in
the figures by the red points.

In the second stage, we extrapolate the corrected points
from the 3 values of $m_y$ to $m_s^{\rm phys}$. 
The dependence here is analytic,
and we use both linear and quadratic forms in these ``Y-fits''.
The Y-fits are straightforward, and we refer to Ref.~\cite{ref:wlee-2010-1}
for examples. Here we focus on the X-fits.
\begin{table}[h!]
\begin{center}
\begin{tabular}{l | c | c | c | c}
\hline
\hline
ID (meas) & $d_0$ & $d_1$ & $d_2$ &$\chi^2/\text{d.o.f}$ \\
\hline
C3 (671 $\times$ 9) & 0.5602(34) & 0.035(16) & ---& 0.83(54) \\
S1 (744 $\times$2) & 0.4808(50) & 0.143(30) & ---& 0.06(14) \\
\hline
S1 (513)  & 0.469(15)  & 0.33(21) & -0.72(81) &  0.002(18) \\
S1 (744 $\times$ 2) & 0.4834(93) & 0.09(13) &  0.27(51) &  0.075(94) \\
\hline
\end{tabular}
\end{center}
\caption{Parameters of X-fits shown in Figs.~\protect\ref{fig:su2-4x3y-nlo}
and ~\protect\ref{fig:su2-4x3y-nnlo}.}
\label{tab:X-fit:C3+S1}
\end{table}

The parameters of the fits shown 
in Fig.~\ref{fig:su2-4x3y-nlo} are given
in the first two rows of Table~\ref{tab:X-fit:C3+S1}.\footnote{%
For a given ensemble with fixed $a m_\ell$, $d_3$ is
absorbed into $d_0$.}
We use uncorrelated
fits and thus expect $\chi^2/{\rm d.o.f}\ll 1$ for a good fit.
The C3 data are unchanged from last year
(Refs.~\cite{ref:wlee-2009-1,ref:wlee-2010-1}) 
while the S1 results have significantly improved statistics.
The curvature in these NLO fit functions is entirely due to the
chiral logarithms, and is consistent with our data on all
ensembles. The convergence of the chiral expansion can be
gauged from the difference between the values of $d_0$ in the
Table (which is the LO result) and the results in the figures.
The ratio of NLO to LO terms is $< 10\%$ for all points used in the
fits. Such satisfactory convergence is 
seen on all other ensembles~\cite{ref:wlee-2010-1}.


The rather large $\chi^2$ in the C3 fit may
be indicative of the need to include NNLO terms,
and indeed the $\chi^2$ of our NNLO fit is much reduced.
The resulting extrapolated-corrected value of $B_K$ (the red point)
is, however, very similar in both fits.

\begin{figure}[t!]
\centering
\includegraphics[width=0.49\textwidth]
{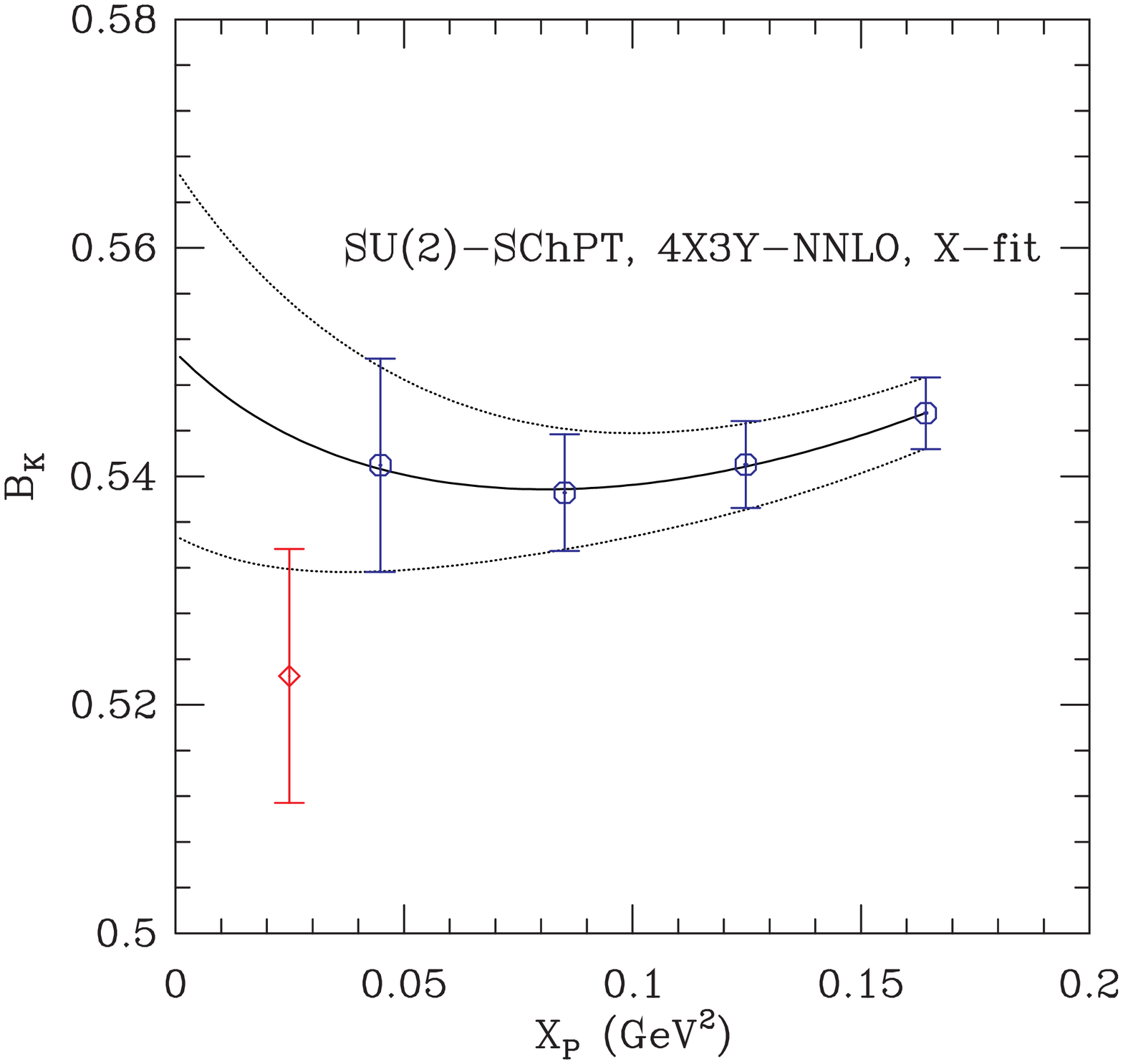}
\includegraphics[width=0.49\textwidth]
{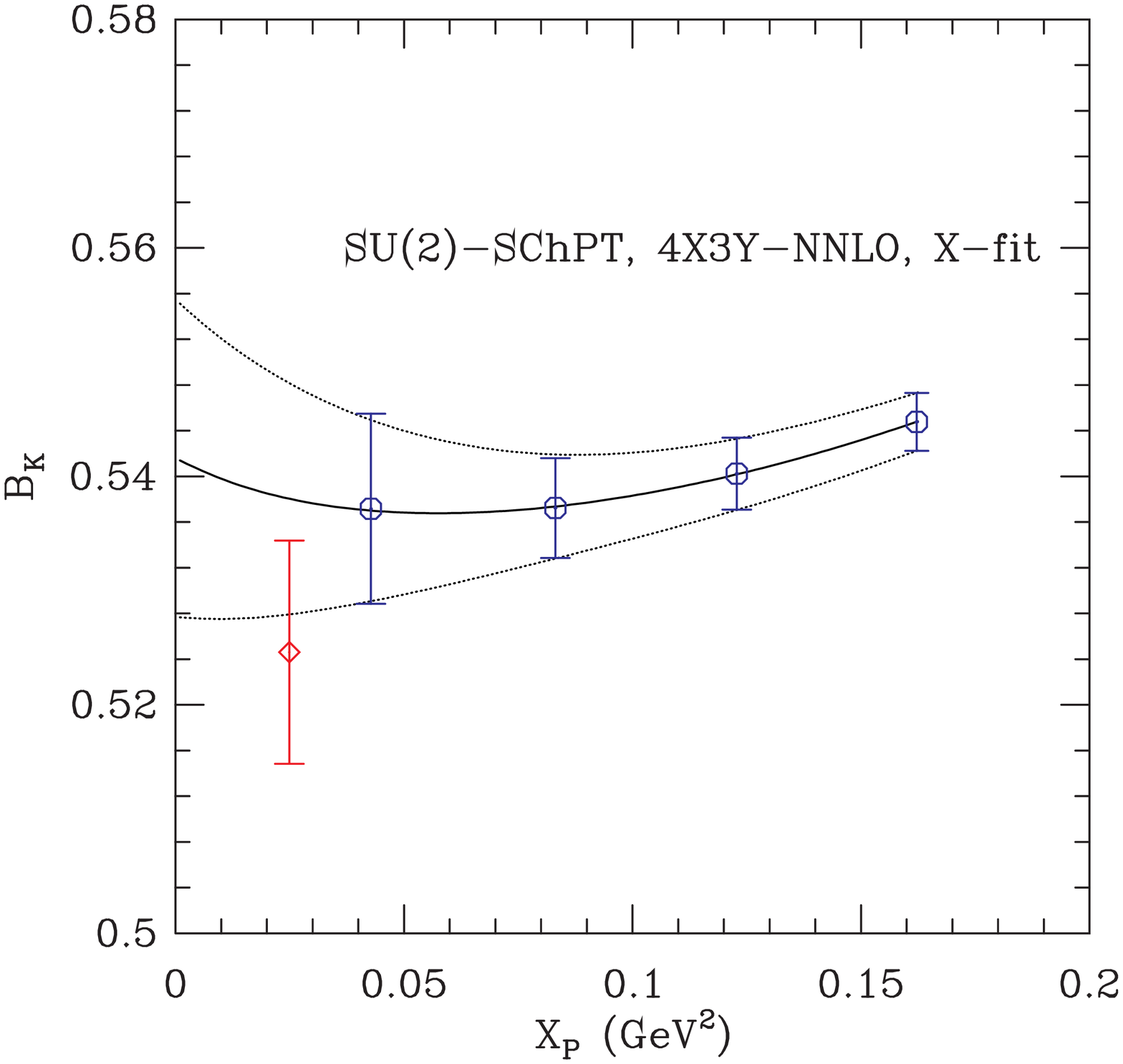}
\caption{$B_K$ versus $X_P$ on the F1 (left) and F2 (right) ensembles,
  showing a 4X3Y-NNLO fit.  Notation as in
  Fig.~\protect\ref{fig:su2-4x3y-nlo}.  Here, we set $am_y = 0.031$
  for both F1 and F2.}
\label{fig:su2-4x3y-nnlo-fine}
\end{figure}

In Fig.~\ref{fig:su2-4x3y-nnlo-fine}, we compare 
4X-NLO fits on the F1 and F2 lattices. The former results are
from last year, while the latter are new.
These two ensembles differ only in the value of light sea-quark mass, and
we see that results are very similar on the two ensembles.
This can be seen quantitatively from the resulting values of
$B_K(2\;{\rm GeV})$ given in Table~\ref{tab:milc-lat}.
The lack of dependence on $a m_\ell$ confirms the result found
on the coarse ensembles C1-C5.
%
%

\begin{figure}[t!]
\centering
    \includegraphics[width=0.49\textwidth]
       {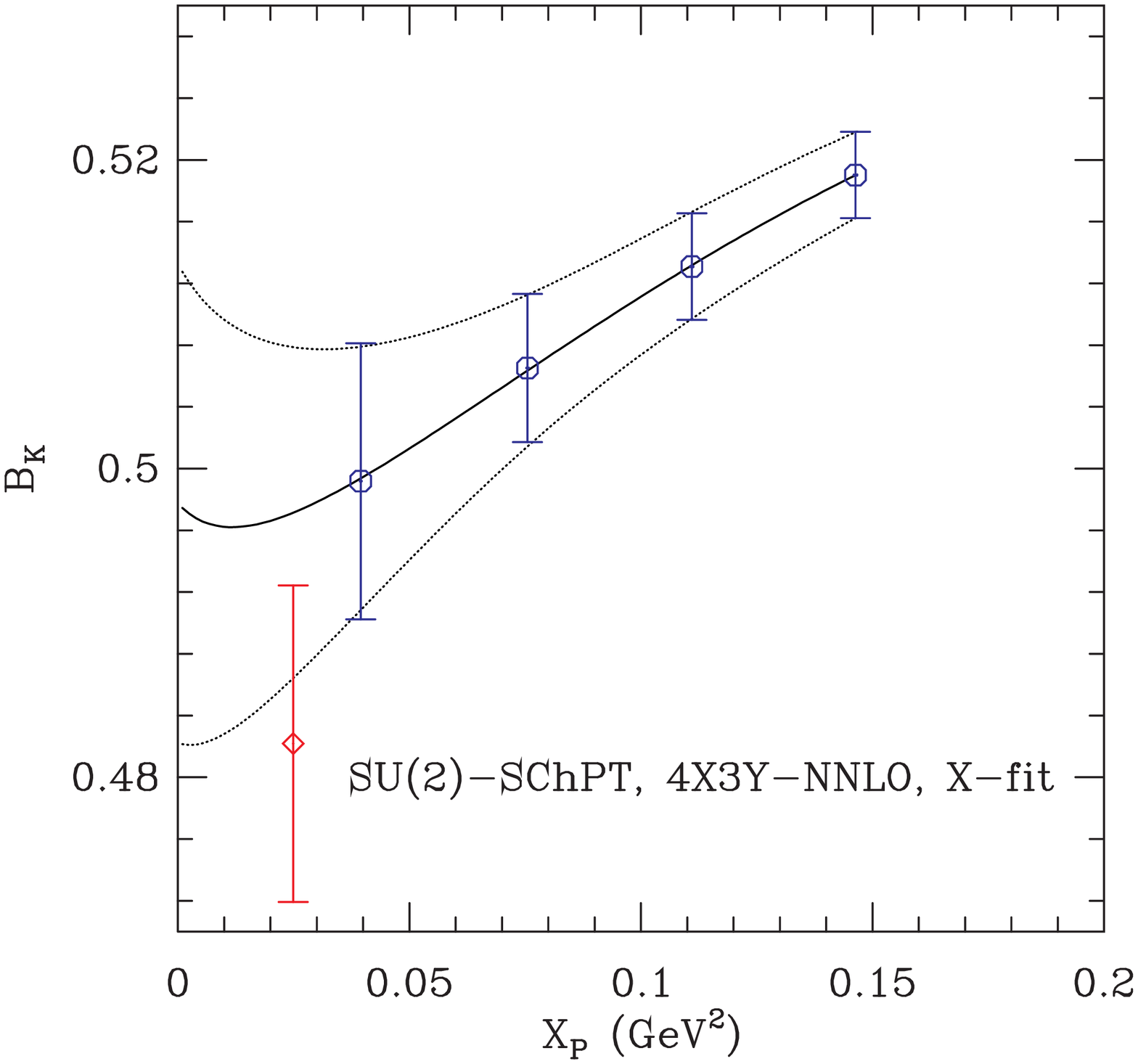}
    \includegraphics[width=0.49\textwidth]
       {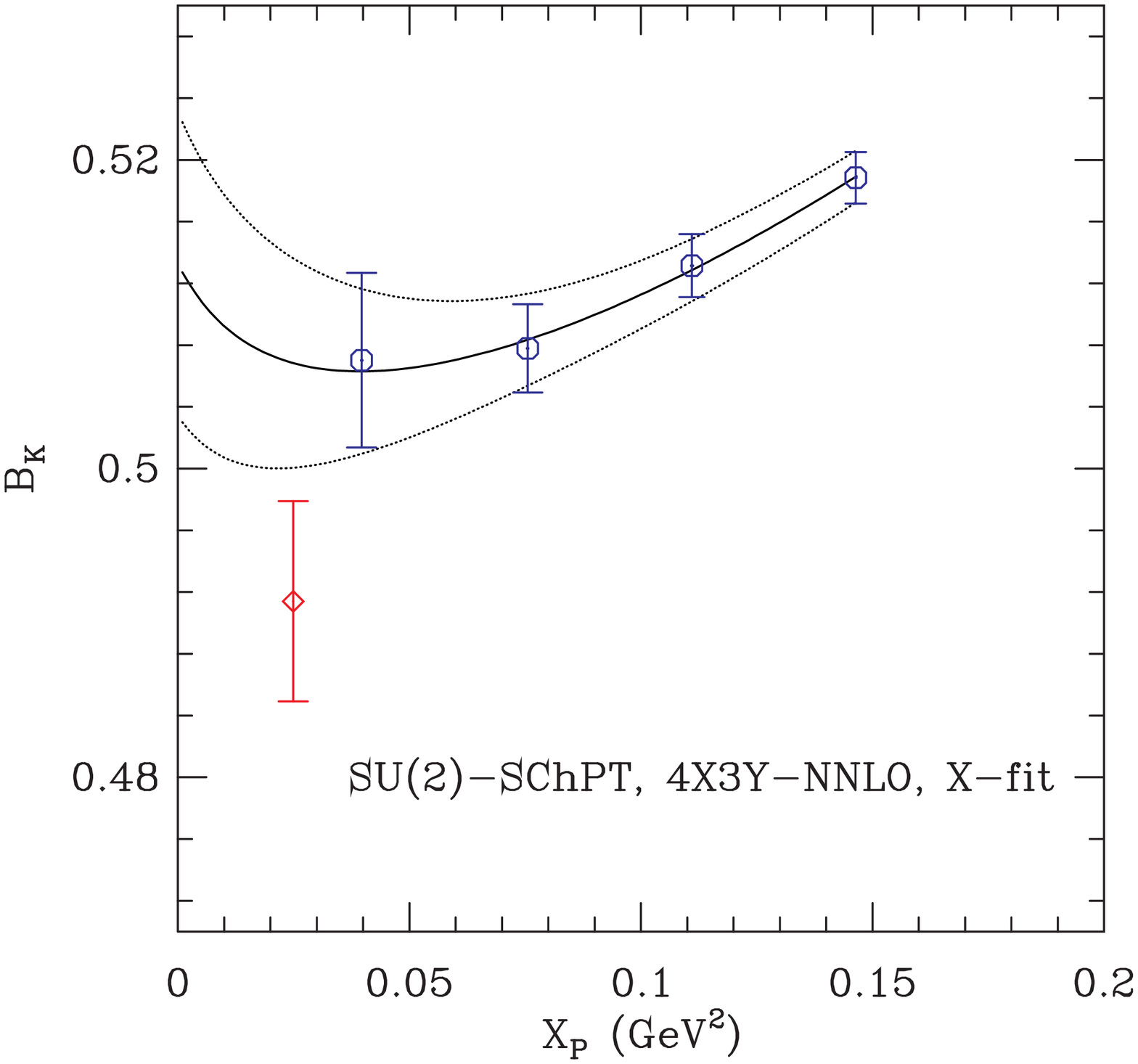}
\caption{$B_K$ versus $X_P$ on S1 ensemble for 513 configurations 
  (left) and for 744 configurations $\times$ 2 measurements (right). 
  The plots are for $am_y=0.018$. 4X3Y-NNLO fits are shown.
  Notation as in Fig.~\protect\ref{fig:su2-4x3y-nlo}.
}
\label{fig:su2-4x3y-nnlo}
\end{figure}

In Fig.~\ref{fig:su2-4x3y-nnlo}, we show the impact on X-fits
of improving the statistics on the S1 ensemble. We compare 
4X3Y-NNLO fits between last year's statistics 
and our new results. 
The corresponding fit parameters are given in Table~\ref{tab:X-fit:C3+S1}.
%
%

%
%

Increasing the statistics makes the visual evidence of
curvature more convincing and improves the determinations
of the LECs.
Comparing the S1 results from NLO fits 
and NNLO fits we see 
from Table~\ref{tab:X-fit:C3+S1})
that $d_0$ and $d_1$ are consistent, although the
latter is poorly determined in both fits.
Fortunately, it is $d_0$ which
gives the dominant contribution to
the physical $B_K$,
and this is well determined  by both fits.

\section{Continuum Extrapolation}
Repeating the above procedure on all ensembles
yields the results for $B_K$ quoted in Table~\ref{tab:milc-lat}.
Here we have run these results 
to a common renormalization scale, $\mu=2\;$GeV.
The next steps are to extrapolate to the continuum limit,
and to estimate all sources of error.

We do the continuum extrapolation using the
results from the C3, F1, S1 and U1 lattices. 
These lattices all have the same ratio of sea quark masses, 
$m_l/m_s = 1/5$ and all have $m_s\approx m_s^{\rm phys}$.
We have seen above that $B_K$ is almost independent of the sea-quark mass,
so the lack of exact matching of the sea-quark masses 
between these ensembles
has a very small effect (and can be corrected for).

The expected approach to the continuum limit is somewhat
complicated.
The dominant errors remaining in $B_K$ are
due to taste-conserving discretization errors and errors
from the truncation of matching factors.
The discretization errors have the form $a^2\alpha_s(1/a)^n$, 
with $n=0,1,2,\dots$.
Note that $n=0$ is allowed because we do not Symanzik-improve our
operators.
%
The truncation errors, by contrast,
 are proportional to $\alpha_s(1/a)^m$ with $m=2,3,\dots$.

\begin{figure}[t!]
\centering
\includegraphics[width=0.5\textwidth]{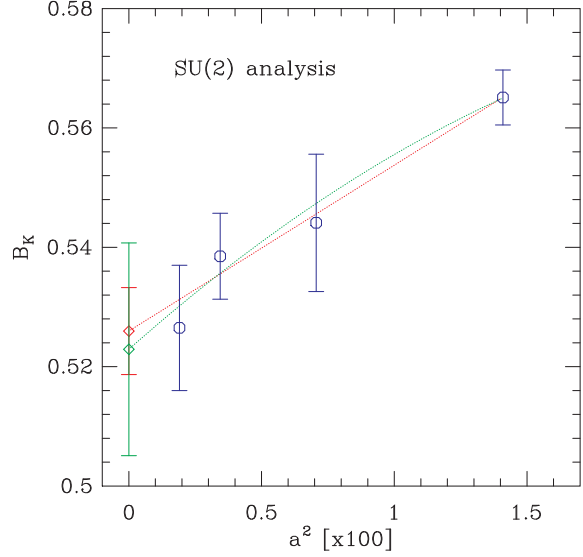}
\caption{Continuum extrapolation of 
$B_K(\text{NDR}, \mu=2\text{ GeV})$. Linear and quadratic
fits to $a^2$ (in fm${}^2\times 100$)  are shown.
The results are from 4X3Y-NNLO SU(2) SChPT fits.}
\label{fig:bk:a^2}
\end{figure}

Since we cannot hope to disentangle these various dependencies,
we adopt a pragmatic approach. We fit to the form
\begin{equation}
B_K(a) = B_K(a=0) + b_1 a^2 + b_2 a^4\,,
\end{equation}
with or without the quadratic term. This takes care of the dominant
$a^2$ term, and the inclusion of the $b_2$ is an approximate way of
allowing for dependence such as $a^2\alpha_s$.
Clearly this fit does not treat the truncation error correctly.
Instead,
we treat this as a systematic error, as discussed below.
In Fig.~\ref{fig:bk:a^2}, we show 
linear and quadratic fits to the $a^2$
dependence of $B_K$.
The fits agree well, and we choose the
linear fit for our central value and statistical error.
We quote the difference between the result on the U1 lattice and
the continuum extrapolated value as an estimate of systematic
error in the continuum extrapolation (due to our not
accounting for $a^2\alpha_s$ terms).

\section{Error Budget and Conclusion}
%
%
%
%
\begin{table}[h]
\centering
\begin{tabular}{ l | l l r }
\hline \hline
cause & error (\%) & memo & status \\
\hline
statistics      & 1.4    & 4X3Y-NNLO fit  & update\\
matching factor & 4.4    & $\Delta B_K^{(2)}$ (U1) & update\\
discretization  & 0.10   & diff.~of (U1) and $a=0$ & update \\
fitting (1)     & 0.92   & X-fit (C3) & \cite{ref:wlee-2010-1}\\
fitting (2)     & 0.08   & Y-fit (C3) & \cite{ref:wlee-2010-1}\\
$am_l$ extrap   & 0.48   & diff.~of (C3) and linear extrap 
                                      & \cite{ref:wlee-2010-1}\\
$am_s$ extrap   & 0.5    & constant vs linear extrap 
                                      & \cite{ref:wlee-2010-1}\\
finite volume   & 0.85   & diff.~of $20^3$ (C3) and $28^3$ (C3-2) 
                                      & \cite{ref:wlee-2010-1}\\
$r_1$           & 0.14   & $r_1$ error propagation 
                                      & \cite{ref:wlee-2010-1}\\
\hline \hline
\end{tabular}
\caption{Error budget for $B_K$ obtained using 
  SU(2) SChPT fitting.
  \label{tab:su2-err-budget}}
\end{table}
The error budget for the SU(2) analysis is presented in Table 
\ref{tab:su2-err-budget}. Several of the errors are as in
Ref.~\cite{ref:wlee-2010-1}, and we refer to that work
for explanations and discussion of these errors.\footnote{%
See also the companion proceedings~\cite{ref:wlee-2010-4} for
additional discussion of errors due to finite-volume effects
and the dependence on sea-quark masses.}
The errors which have changed since Ref.~\cite{ref:wlee-2010-1}
are those due to 
statistics (reduced from 1.7\%), 
matching (reduced from 5.5\%) and
discretization (reduced from 1.8\%).
Our estimate of the latter error,
explained in the previous section, may be an underestimate,
but in any case is dominated by the matching error.

The matching error is estimated as follows.
We assume that the dominant missing term in the perturbative 
matching factors is of size $1 \times \alpha_s^2(\mu=1/a)$,
so that the error in $B_K$ is
\begin{equation}
\Delta B_K^{(2)} \approx B_K^{(1)} \times \Big[\alpha_s(1/a)\Big]^2\,,
\end{equation}
where $B_K^{(1)}$ is the result using one-loop matching.
This error will be reduced, but not eliminated,
by the continuum extrapolation.
To be conservative, 
we take as the error the size of $\Delta B_K^{(2)}$ 
on our smallest lattice spacing.
The reduction in this error compared to Ref.~\cite{ref:wlee-2010-1}
is simply due to our having another, smaller, lattice spacing.

Combining the errors in the error budget, we find
\begin{equation}
\begin{array}{l l}
 B_K(\text{NDR}, \mu = 2 \text{ GeV}) & = 0.5260 \pm 0.0073 \pm 0.0244 
 \\
 \hat{B}_K = B_K(\text{RGI}) & = 0.720 \pm 0.010 \pm 0.033\,.
\end{array}
\end{equation}
Our result is in agreement with those obtained using
valence DWF on either MILC~\cite{ALV}  
or DWF lattices~\cite{su2chpt2,DWFBK}.
Our total error of 5\% is somewhat larger than the 4.1\% and
3.3\% attained in the other calculations, the difference being
mainly due to our use of one-loop, rather than non-perturbative,
matching.

\section{Acknowledgments}
C.~Jung is supported by the US DOE under contract DE-AC02-98CH10886.
The research of W.~Lee is supported by the Creative Research
Initiatives Program (3348-20090015) of the NRF grant funded by the
Korean government (MEST). 
The work of S.~Sharpe is supported in part by the US DOE grant
no.~DE-FG02-96ER40956.
Computations were carried out in part on QCDOC computing facilities of
the USQCD Collaboration at Brookhaven National Lab. The USQCD
Collaboration are funded by the Office of Science of the
U.S. Department of Energy.

\end{document}